\documentclass[twocolumn,showpacs,preprintnumbers,amsmath,amssymb,superscriptaddress]{revtex4-1}   

\usepackage{graphicx}
\usepackage{dcolumn}
\usepackage{empheq}
\usepackage{amsmath,mathrsfs,amssymb,amsthm}  
\usepackage{wasysym,enumerate,color}
\usepackage{bm}
\definecolor{darkgreen}{rgb}{0,0.5,0}
\definecolor{orange}{rgb}{1,0.5,0}
\definecolor{grey}{rgb}{.6,.6,.6}

\newcommand{\bra}[1]{\langle #1|}
\newcommand{\ket}[1]{|#1\rangle}

\newcommand{\be}{\begin{equation}}
\newcommand{\ee}{\end{equation}}
\newcommand{\beq}{\begin{eqnarray}}
\newcommand{\eeq}{\end{eqnarray}}

\begin{document}

\title{Out-of-time-ordered commutator across an impurity quantum phase transition}
\title{Information scrambling at an impurity quantum critical point}

\author{Bal\'azs D\'ora}
\email{dora@eik.bme.hu}
\affiliation{Department of Theoretical Physics and MTA-BME Lend\"{u}let Spintronics 
Research Group (PROSPIN), Budapest University of Technology and Economics, 1521 Budapest, Hungary}
\author{Mikl\'os Antal Werner}
\affiliation{BME-MTA  Exotic  Quantum  Phases  Research Group,   Budapest  University  of  Technology  and  Economics,  1521  Budapest,  Hungary}
\author{C\u at\u alin Pa\c scu Moca}
\affiliation{BME-MTA  Exotic  Quantum  Phases  Research Group,   Budapest  University  of  Technology  and  Economics,  1521  Budapest,  Hungary}
\affiliation{Department  of  Physics,  University  of  Oradea,  410087,  Oradea,  Romania}

\date{\today}

\begin{abstract}
The two-channel Kondo impurity model realizes a local non-Fermi liquid state with finite residual entropy.
The competition between the two channels drives the system to an impurity quantum critical point.
We show that the out-of-time-ordered (OTO) commutator for the impurity spin 
reveals markedly distinct behaviour depending on the low energy impurity state.
For the one channel Kondo model with Fermi liquid ground state, the  OTO commutator vanishes for late times,  indicating the absence of the butterfly effect.
For the two channel case, the impurity OTO commutator is completely temperature independent and  saturates quickly to its upper bound  1/4, 
and the butterfly effect is maximally enhanced. These compare favourably to numerics on  spin chain representation of the Kondo model. Our results imply that a large late time value of the OTO commutator does not necessarily diagnose quantum chaos.

\end{abstract}


\maketitle

\section{Introduction}

Non-Fermi liquids with finite residual entropy at vanishing temperature are peculiar states of matter, which
hold the promise to be relevant for holography and can constitute the holographic duals of black holes~\cite{sachdevprx}.
Particularly interesting in this context is the Sachdev-Ye-Kitaev model~\cite{sachdevye,maldacenaprd,larkin,fritzpatrick},
describing  randomly interacting Majorana fermions, which could possibly be related to quantum gravity.

In order to diagnose information scrambling and the related quantum butterfly effect in
the Sachdev-Ye-Kitaev model, the
out-of-time-ordered (OTO) commutator~\cite{larkin} was proposed~\cite{sachdevye,maldacenaprd}
as
\begin{gather}
C(t)=-\left\langle \left[V^{},W(t)\right]^2\right\rangle\geq 0.
\label{ct}
\end{gather}
Here, $V$ and $W$ are some local hermitian operators,
and $W(t)=\exp(iHt)\,W\exp(-iHt)$.
Assuming, that the involved $V$ and $W$ operators commute at $t=0$,
the $C(t)$ measures how commutativity is destroyed during the time evolution. 
For a sufficiently
chaotic system, the commutator is argued to exhibit exponential temporal
growth~\cite{roberts,roberts2016}, bounded by a thermal Lyapunov
exponent~\cite{maldacena2016} and to become large in the long time limit, hence the butterfly effect appears.
This occurs through the vanishing of the  $\langle V\,W(t)\,V\,W(t)\rangle$  OTO correlator,
investigated in a variety of systems~\cite{nyao,swingle,zhugrover,campisi,aleiner,tsuji,bohrdt,chowdhury},  and recently measured experimentally~\cite{junli,garttner,meier}.
In systems displaying Fermi liquid behavior~\cite{roberts2016,tsuji,doramoessner}, the large time limit  of $C(t)$ seems to approach zero,
a behaviour which is associated to the presence
of the fermionic excitations at the Fermi surface.
On the other hand, the quasiparticle picture is lost in a non-Fermi liquid, 
and understanding the behaviour of $C(t)$ is important in such exotic states.

Non-Fermi liquid phases arise typically in strongly interacting models, where enhanced quantum fluctuations destroy fermionic excitations and give way
to collective modes~\cite{giamarchi,nersesyan,herbut2014}.
Among the quantum impurity models, the  two channel Kondo (2CK) model, which has already been realized  experimentally~\cite{potok,iftikhar,keller},
is the most promising candidate as it occurs
at an impurity quantum critical point (iQCP)  and displays a non-Fermi liquid behavior. In spite of being a traditional condensed matter model, its holographic realization has already attracted attention~\cite{Erdmenger2013}.

In contrast to the one channel Kondo (1CK) problem,
where the ground state becomes a Fermi liquid
and the impurity spin is completely screened at low temperatures~\cite{hewson},
in the two channel version the physics is drastically different.
When the two channels are symmetrically coupled to the
impurity, the conduction electrons within each channel compete without success to screen the impurity.
This frustration creates a non-Fermi liquid ground state, reflected in a $\ln{2}/2$ residual entropy 
as half of the impurity degrees of freedom are completely decoupled~\cite{emery}.
The 2CK behaviour occurs at an iQCP between two 1CK regions, whose
possible order parameter has attracted revived interest quite recently~\cite{bayat,Alkurtass}.

Therefore, it looks relevant to address the behaviour of
 the OTO commutator in the anisotropic 2CK model, realizing both 1CK and 2CK physics with the hope of disentangling the effect of
the Fermi vs. non-Fermi liquid character of the ground states on $C(t)$. 
We find that although the simple commutator of the impurity spin \`a la Kubo formula is temperature dependent, the OTO commutator
turns out to be completely temperature independent for the perfect 2CK model.
While commutativity in the OTO  commutator is restored at late times for the 1CK case,
the 2CK model features a maximally enhanced late time value $C(t\to \infty)= 1/4$ and a maximal quantum butterfly effect. As we show,
this occurs due to the decoupled Majorana mode in the 2CK case.
We emphasize that unlike in chaotic models, where the OTO correlator is expected to vanish~\cite{roberts},
for the 2CK model this correlator changes sign with respect to its $t=0$ value during the time evolution,
and the late time value of the OTO commutator is twice as big as expected in suitably chaotic systems,
even though the 2CK model is integrable and not chaotic.

\section{The Kondo impurity model}

The two-channel Kondo impurity model Hamiltonian~\cite{hewson}  is given by 
\begin{gather}
H_K=\sum\limits_{j=1}^2\left\{ \sum_{p,s}\epsilon(p)c^+_{p,s,j}c_{p,s,j}+\sum_{\gamma=x,y,z}J_{\gamma,j} s_\gamma S_{\gamma,j}(0)\right\},
 \label{Hk}
\end{gather}
where $S_{\gamma,j}(0)=\sum_{s,s'} \Psi_{s,j}^+(0)\sigma_{s,s'}^\gamma\Psi_{s',j}(0)$, $s$ denotes the spin quantum number of the conduction electrons, 
$\epsilon(p)$ is their kinetic energy,
the $\sigma$'s are Pauli matrices,
 $s_{x,y,z}$ stands for the impurity spin components, $c_{k,s,j}$ and $\Psi_{s,j}(0)$ are the conduction electron annihilation operators with spin $s$ and channel $j$
 in momentum and real space, respectively.
In addition, we require XXZ couplings as $J_{x,j}=J_{y,j}$ and $J_{z,1}=J_{z,2}$.
In case of channel isotropy, i.e. $J_{x,1}=J_{x,2}$, 
Eq. \eqref{Hk} realizes the 2CK model, 
otherwise the low energy physics is governed by the one channel case, and a crossover to the two-channel behaviour can occur with
increasing energy, unless the model is completely anisotropic.

Upon Abelian bosonization~\cite{giamarchi,nersesyan}, this problem can be mapped onto the Majorana resonant level (MRL) 
model~\cite{emery}. This reads as
\begin{gather}
H_{mrl}=H_0(\xi)+H_0(\eta)-i\frac{I_+}{\sqrt{2\pi\alpha}}a\,\xi(0)-i\frac{I_-}{\sqrt{2\pi\alpha}}b\,\eta(0),
\label{Hmrl}
\end{gather}
where $a=a^+$, $b=b^+$ are the impurity Majorana operators with $a^2=b^2=1/2$, $\xi(x)$ and $\eta(x)$ stem 
from the Majorana representation of conduction electrons~\cite{nersesyan},
$H_0(\zeta)= iv\int dx\zeta(x)\partial_x\zeta(x)$ with $v$ the Fermi velocity, 
$I_\pm=(J_{x,1}\pm J_{x,2})/2$ measures the channel anisotropy and $J_{z,1}=2\pi v$ at the 
Emery-Kivelson~\cite{emery} or Toulouse~\cite{nersesyan} 
point, where the above MRL model description holds, $\alpha$ plays the role of the remnant of the lattice constant in the low energy theory, and its
inverse serves as a high energy cutoff.
Note that Eq. \eqref{Hmrl} contains a linearized dispersion for the conduction electrons, and neglects curvature effects, as is customary in similar approaches~\cite{giamarchi}.

The $(\xi,a)$ part of the Hamiltonian commutes with the $(\eta,b)$ sector, therefore the dynamics of the two
Majoranas decouples and can be considered separately.
The $I_-=0$ marks the iQCP, and the low energy physics is equivalent to 2CK, while for
$I_+=I_-$, the low energy dynamics of the 1CK case is realized. In between, a crossover from one to 
two-channel behaviour takes place with increasing energy,
and $I_-<I_+$. 
The $z$ component of the impurity spin, what we are going to investigate here, is $s_z=iab$, while the other spin components are more complicated due to the involved unitary transformations in the mapping from Eq. \eqref{Hk}
to Eq. \eqref{Hmrl} via Refs. \cite{emery,giamarchi,nersesyan}.


\section{Majorana propagators} 

\begin{figure}[h!]
\includegraphics[width=0.75\columnwidth]{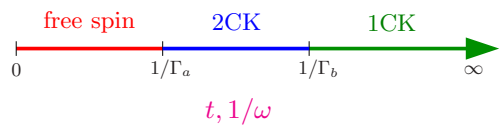}
\caption{Schematic "phase diagram" of the anisotropic two channel Kondo model at zero temperature
in the time or energy domain. Only for $\Gamma_b=0$ is the low energy physics
governed by the two channel Kondo effect.}
\label{kondophasediag}
\end{figure}

Since the MRL model is quadratic, the Matsubara Green's function of the local Majorana operators are evaluated as
$G_{a,b}^{-1}(i\omega_n)={i\omega_n+i\Gamma_{a,b}\textmd{sign}(\omega_n)}$,
 where $\omega_n$ is the fermionic Matsubara frequency, and $\Gamma_{a,b}=I_\pm^2/(4\pi v\alpha)$, respectively.
The schematic phase diagram of the anisotropic 2CK model with the crossover regions is depicted in Fig. \ref{kondophasediag}.
At short times or high energies, the spin is practically not influenced by the electrons, and behaves as a free spin.
With increasing time/decreasing energy, we enter into the 2CK regime, unless the model
is completely anisotropic and corresponds to 1CK.
The late time behaviour is of 2CK type only in case of perfect isotropic couplings, otherwise
it is dominated by 1CK physics.

We emphasize that the MRL model in Eq. \eqref{Hmrl}  corresponds to the strong coupling limit of the Kondo problem, when the exchange coupling, $J_{z,1}$ 
is of the order of the electronic
bandwidth. Consequently, $\Gamma_{a,b}$ is not the Kondo temperature but rather describes how fast we move along the Emery-Kivelson line towards the infinitely strong coupling fixed point~\cite{nersesyan}.
As noted above, at the Emery-Kivelson point, $I_-=0$ and one of the Majorana modes decouples  completely at the iQCP, 
and its Green's function becomes $1/i\omega_n$.

In the following, we will need the propagator of the Majorana fermions in real time. We follow Ref. \cite{zubarev}
to obtain
\begin{gather}
F_a(t)=\langle a(t)a \rangle=\int\limits_{-\infty}^\infty d\omega \frac{\rho_a(\omega)}{\exp(\omega/T)+1}\exp(-i\omega t),
\label{fat}
\end{gather}
and $\rho_a(\omega)=-\textmd{Im}G_a(i\omega_n\rightarrow \omega+i0^+)/\pi=\Gamma_a/(\omega^2+\Gamma_a^2)\pi$ is the density of states, $T$ is temperature and 
similar expression holds for the $b$ Majorana fermion with $\Gamma_a\rightarrow\Gamma_b$ change.
Since $\Gamma_b<\Gamma_a$, it is the $b$ Majorana field which decouples completely from conduction electron when $I_-=0$, in which case its
density of states becomes a Dirac-delta function as $\rho_b(\omega)=\delta(\omega)$, and $F_b(t)\equiv 1/2$.

At $t=0$, $F_{a,b}(0)=1/2$, in accord with the definition of Majorana operators. 
At $T=0$, the above integral is performed analytically to yield
\begin{gather}
F_a(t)=\frac 12\exp(-\Gamma_at)+\frac{i}{2\pi}\sum_{s=\pm}s\exp(-s\Gamma_at)\textmd{Ei}(s\Gamma_at),
\label{majoranaprop}
\end{gather}
where Ei$(x)$ is the exponential integral function~\cite{gradstein}. For late times, the propagator decays as
$F_a(t\gg 1/\Gamma_a)=i/(\pi\Gamma_a t)$.
It is important to note that temperature influences only the imaginary part of the Majorana propagator, the real part remains temperature independent, i.e.
Re$F_{a,b}(t)=\exp(-\Gamma_{a,b}t)/2$ for all temperatures.
Its imaginary part acquires an additional $\exp(-\pi Tt)$ factor to the power law decaying part.

\section{Analytical results for the simple and the OTO commutators of the impurity spin}
\label{sec:analytics}

Let us start with the simple commutator of the impurity spin, whose Fourier transform is the dynamic spin susceptibility~\cite{emery}, $\chi(\omega)$, accessible by
e.g. electron spin resonance or neutron spectroscopy.
It is given by
\begin{gather}
K(t)=i\langle\left[ s_z(t),s_z\right]\rangle=2\textmd{Im}\left(F_a(t)F_b(t)\right).
\label{k(t)}
\end{gather}
For the 2CK case at $T=0$, $F_b(t)=1/2$, therefore it decays in a power law fashion for long times as 
$K(t\gg 1/\Gamma_a)\sim 2/\pi\Gamma_a t$, which translates to a dynamic
spin susceptibility as Im$\chi_{2CK}(\omega)\sim\textmd{sign}(\omega)$ at low frequencies~\cite{emery}. 
Away from the two channel point, the low energy physics is determined by the 1CK effect, therefore 
$K(t\gg 1/\Gamma_b)\sim \exp(-\Gamma_b t)/\pi\Gamma_a t$, giving
rise to a spin susceptibility 
Im$\chi_{1CK}(\omega)\sim\omega$  at low energies~\cite{rmpleggett}.
For $1/\Gamma_a\ll t \ll 1/\Gamma_b$, on the other hand, the 2CK behaviour is recovered.
In the isotropic situation with $\Gamma_a=\Gamma_b$,  there is no 
crossover towards the two  channel behavior, and the decay for $t\gg 1/\Gamma_a$ is
$K(t)\sim 2\exp(-\Gamma_a t)/\pi\Gamma_a t$. 
The hierarchy of energy scales is clearly observable, and depending
on the anisotropies, there can be a wide enough temporal window to catch 2CK in the act.
At finite and large enough temperatures, i.e. $T>\Gamma_{a,b}$, the damping factor $\Gamma_{a,b}$ in the exponents is replaced by the temperature.
The Emery-Kivelson point thus reproduces the generic behaviour of the Kondo model for $K(t)$, as we also show below in Fig. \ref{fig:K}.

The OTO commutator for the impurity spin reads as
\begin{gather}
C(t)=-\langle\left[ s_z(t),s_z\right]^2\rangle,
\end{gather}
which, based on the Cauchy-Schwarz inequality, is bounded from above for a spin-1/2 impurity as
$C(t)\le 4 \| s_z\|^4=1/4$.
Using the fact that $s_z^2=1/4$, the OTO commutator simplifies to
\begin{gather}
C(t)=2\left(\frac{1}{2^4}-\textmd{Re}f(t)\right),
\label{otockondo}
\end{gather}
where the OTO correlator is defined as
\begin{gather}
f(t)=\langle s_z(t)s_zs_z(t)s_z\rangle.
\label{otocorrelator}
\end{gather}
Keeping in mind that $s_z = i a b$ and the fact that $(\xi, a)$ and $(\eta,b)$ sectors commute 
the OTO correlator can be factorized as
\begin{gather}
f(t)=\langle a(t)a a(t)a\rangle \langle b(t)b b(t)b\rangle.
\label{otocorrelator1}
\end{gather}
Then, since MRL Hamiltonian in Eq.~\eqref{Hmrl} is quadratic, its ground state is a Slater determinant. For such a state, we use the decoupling in each sector. In the $(\xi,a)$ sector 
we have~\cite{fn3}
\begin{eqnarray}
\langle  a(t)a a(t)a \rangle&=& \langle  a(t)a \rangle ^2 -\langle a(t)a(t)\rangle \langle a a\rangle
+\langle  a(t)a \rangle \langle  a a(t) \rangle\nonumber \\
	&=& F_a(t)^2-\frac 14+\left|F_a(t)\right|^2\\
	& =&2F_a(t)\textmd{Re}F_a(t)-\frac 14\nonumber
\end{eqnarray}
where we have used that $F_a(-t)=F^*_a(t)$ and $a^2=a(t)^2=1/2$.
Similar considerations apply also to the $b(t)$ Majorana operators. 
Then the OTO correlator is expressed exactly as 
\begin{gather}
f(t)=\prod_{q=a,b}\left(2F_q(t)\textmd{Re}F_q(t)-\frac{1}{4}\right).
\end{gather}
The OTO correlator starts from $1/16$ at $t=0$ and recovers its initial value for 1CK at late times while
for the 2CK, it approaches $-1/16$.  

\begin{figure}[h!]
\includegraphics[width=0.85\columnwidth]{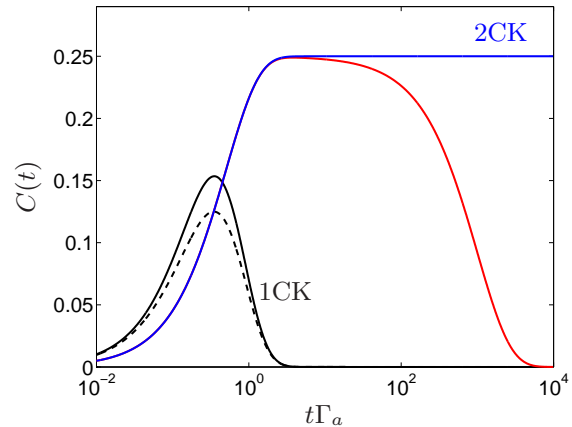}
\caption{The $s_z$ OTO commutator is shown in the perfect one- and two channel cases at $T=0$, $\Gamma_a=\Gamma_b$ (black)
and $\Gamma_b=0$ (blue), respectively, exhibiting also the crossover for $\Gamma_b=\Gamma_a/2000$ (red).
The dashed line denotes $T\gg\Gamma_a$ for the channel isotropic 1CK, when the imaginary part of the Majorana propagator vanishes.
 The perfect 2CK is $T$ independent while for the crossover case, all finite $T$ data falls on top of the zero temperature curve.}
\label{szotoc}
\end{figure}

Having the formalism implemented, we can now compute the OTO commutator. For the 2CK, it depends only on the real part of $F_a(t)$ from Eq. \eqref{majoranaprop} since
$F_b(t)\equiv 1/2$ is purely real, hence it is completely independent of the temperature.
Although the iQCP is located at $T=0$, its makes its presence felt also in the finite temperature response\cite{sachdev}. 
The OTO commutator is
evaluated at arbitrary times as
\begin{gather}
C_{2CK}(t)=\frac{1-\exp(-2\Gamma_at)}{4}\xrightarrow{ t\rightarrow\infty }\frac 14,
\label{otoc2ck}
\end{gather}
saturating to a finite value at late times.
For the one channel case with general anisotropies, the OTO commutator exhibits very mild temperature dependence (see Fig. \ref{szotoc}), albeit 
 at late times it becomes temperature independent and vanishes as
\begin{gather}
C_{1CK}(t\rightarrow\infty)\sim \frac{\exp(-2\Gamma_{b}t)}{4}.
\label{eq:C_1CK}
\end{gather}
This agrees with the late time limit of Eq. \eqref{otoc2ck} in the 2CK limit, when $\Gamma_b=0$.
Note that the temperature independence of the OTO commutator  appears to be a generic feature 
in a non-Fermi liquid ground state, as it was also found in Luttinger liquids as well~\cite{doramoessner}.
Eq.~\eqref{otockondo} predicts an initial  linear increase in time,
though Eq.~\eqref{ct} only allows for a quadratic time dependence,
 since a high energy cutoff, accounting for the bandwidth of conduction electrons in 
 Eq.~\eqref{Hk}, was sent to infinity.
Had we retained this cutoff, the correct $t^2$ increase would be recovered.
For the two channel case at late times, i.e. $t\gg1/\Gamma_a$, the OTO commutator reaches it maximal possible value:
it consists of 4 terms, each of which is bounded from above by $1/16$, since they all contain 4 spin-1/2 operators.
Each term takes on its maximal value, therefore $C(t)\rightarrow 1/4$ rapidly with increasing time for 2CK, though the model itself
is integrable. 
The maximal value of the squared commutator implies that in the long time limit, the squared anticommutator vanishes, i.e.
$\langle\left\{ s_z(t),s_z\right\}^2\rangle\xrightarrow{ t\rightarrow\infty }0$.
 Note that in a suitably chaotic system,
the OTO commutator is expected to saturate to $2/2^4=1/8$, exactly half of the value for the 2CK case.
These features are shown in Fig. \ref{szotoc} for zero and very high temperatures. For intermediate temperatures, the OTO commutator
takes a value between the two extreme temperature limits.

In terms of information scrambling, the local Fermi liquid ground state of the 1CK behaves similarly to that in a Fermi gas, namely that the initially zero
value of the commutator is recovered at late times. As opposed to that, the 2CK case qualifies as a slow information scrambler due to the initial power-law increase of the OTO
commutator, while the large late time value parallels to the one found in another non-Fermi liquid system, in a Luttinger liquid~\cite{doramoessner}
The scrambling time, i.e. the time at which the enhancement occurs, is given by $1/\Gamma_a$.

\section{Numerics}

In order to test the validity of the analytical predictions discussed in Sec.~\ref{sec:analytics},
at the Emery-Kivelson point one can either use perturbation theory by 
 considering additional marginal terms in the 
Hamiltonian~\cite{fn1},
which are e.g. responsible for the logarithmic corrections $\sim T\ln(T)$ for the specific heat in the 2CK case~\cite{emery},
or one can resort to numerics where all these additional processes are taken into account exactly. 
We have decided to follow the latter option, therefore, 
in this section we test numerically the predictions for the OTO commutators  discussed in Sec.~\ref{sec:analytics}.

\begin{figure}[h!]
\includegraphics[width=0.95\columnwidth]{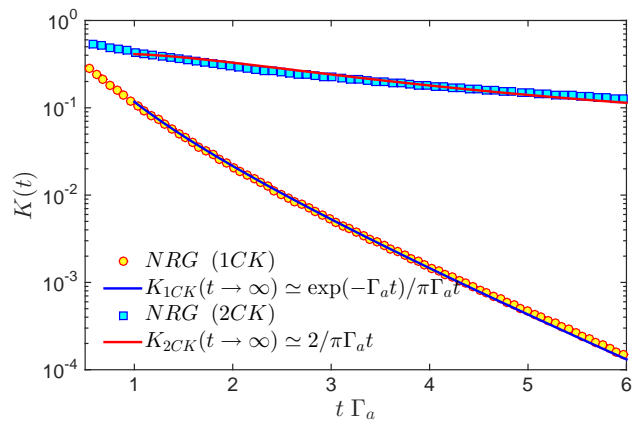}
\caption{ Comparison of the long time behavior $t\gtrsim 1/\Gamma_a$ for the  $K(t\to\infty)$ and the NRG results obtained by Fourier transform  the spin susceptibility.}
\label{fig:K}
\end{figure}

To start with, we have investigated the long time dynamics of $K(t)$ using the numerical
renormalization group (NRG) approach~\cite{Wilson.75} which is the method of choice 
in quantum impurity problems. 
For that, we have computed the spin $s_z$ susceptibility $\chi(\omega)$ and then took the Fourier transform
to get the temporal dependence in $K(t)$. It confirms  that even away from the Emery-Kivelson point,
our temporal scaling discussed in Sec~\ref{sec:analytics} remains intact.
A typical comparison between the analytical results and the NRG data for 
$K(t)$, in the long time limit $t > 1/\Gamma_a$
is displayed in Fig.~\ref{fig:K} which clearly shows the strong $\sim \exp(-\Gamma_a t)/\Gamma_a t$ decay
for the 1CK and the power like $\sim 1/\Gamma_a t$ decay for the 2CK problem. 

On the other hand, the NRG calculation of the OTO correlator would require besides 
the matrix elements of the $s_z$ operator 
between the ground state and an excited state, i.e. $\langle 0|s_z|n\rangle$,
the  matrix elements between two arbitrary excited states, i.e. $\langle m|s_z|n\rangle$, which
spoils the NRG calculation for $C(t)$. In this way the NRG is no longer suitable to compute the OTO commutator
and we need to rely on other methods such as exact diagonalization  or the time evolving block 
decimation.  These approaches shall be discuss in Sec.~\ref{sec:TEBD}.

\subsection{Spin Kondo model}

The standard Kondo Hamiltonian, i.e. Eq.~\eqref{Hk},  contains the Heisenberg interaction between the impurity spin
and the non-interacting electrons.  The spin Kondo model neglects the charge degrees of freedom but
accounts properly for the spin. In this way it reduces considerably  the size of the Hilbert space, making it more amenable for numerically exact studies on finite chains.
The spin chain representation of the one channel Kondo model  (1CK) was originally introduced in Refs~\cite{Laflorencie, Affleck09}, while its extension to two channels is straightforward~\cite{Alkurtass}.
The 2CK version for the spin Kondo model consists of two open Heisenberg chains coupled to a spin-1/2 impurity, whose  Hamiltonian is~\cite{bayat,Alkurtass}
\begin{gather} 
H=\sum_{m=L,R}\left[
J'_{m}\left(J_1{\bf s}\cdot{\bf S}_m^{1}+J_2{\bf s}\cdot{\bf S}_m^{2}\right)+\right.\nonumber \\
\left.
+J_1\sum_{l=1}^{N_{m}-1}{\bf S}_{m}^l\cdot {\bf S}_{m}^{l+1}+J_2 \sum_{l=1}^{N_{m}-2}{\bf S}_{m}^l\cdot{\bf S}_{m}^{l+2}\right],
\label{spinonly}
\end{gather}
where ${\bf s}$ and ${\bf S}^l_{m}$ represent the impurity spin and the spin at site $l$ in channel $m$, respectively, and $N_m$ is the number of spins in chain $m$ and the total number of spins is $N=N_L+N_R+1$. 
The second and the third terms in Eq.~\eqref{spinonly} contain 
nearest-neighbor ($\sim J_1$) and next-nearest-neighbor ($\sim J_2$) spin interactions
in the two ($L/R$) channels. In the first term,  $j^{(1/2)}_{L/R}=J_{L/R}' J_{1/2}$ are the antiferromagnetic Kondo couplings between the impurity spin $\bf s$ and spins in the two channels. The one channel version 
of the model can be obtained simply by setting $J_R'=0$, i.e. decoupling one lead, in Eq~\eqref{spinonly}.
The Hamiltonian~\eqref{spinonly} with $J_2=0$ is Bethe ansatz integrable with a gapless spectrum.  
As $J_2$ becomes finite, a transition to a dimerized phase takes place at $J_2^{(c)}\approx 0.2412\, J_1$ and the energy spectrum becomes gapped. In the numerical calculation it is then suitable to choose $J_2= J_2^{(c)}$, otherwise 
a marginal coupling arises, producing logarithmic corrections that spoil the numerical data.

Furthermore, it has been shown in Ref.~\cite{Laflorencie} that at the critical 
point $J_2 = J_2^{(c)}$ the usual Kondo physics described by the regular Hamiltonian~\eqref{Hk} is exactly captured by the spin version, i.e. Eq.~\eqref{spinonly}. This observation allows
us to speed-up the computation substantially. In this regard,  the main appeal of using  the spin 
chain representation, i.e. Eq. \eqref{spinonly},  is that while the local impurity Hilbert space has the same size as for Eq. \eqref{Hk},  the Hilbert space for the conduction electrons is severely reduced:
the original model contains two spinful channels, with local Hilbert space dimension of $4\times 4=16$ (each for comes from empty, spin up, spin down and doubly occupied states), while the spin only lattice realization requires
two spinless channels with local Hilbert space dimension of $2\times 2=4$.

In general the nature of the Kondo effect, 
and the Kondo temperature $T_K$ in particular, is controlled by the dimensionless couplings as $J'_{L}=J'_{R}$ for 2CK, while in the 1CK situation only by $J'=J'_{L}$ for example, as $J'_{R}$ is fixed to zero. 
In the 1CK problem, for $J'\leq 1$, $T_K$, which is the typical energy scale for the formation of Kondo effect, 
is large and comparable to $J_1$ (see inset of Fig. \ref{kondoed}). 
We use this observation in our favour as it allows us to follow the time evolution of the 
system for $t\gg 1/T_K$.


\subsection{Time evolving block decimation and Exact diagonalization}
\label{sec:TEBD}

In this section we present details for the calculation of the  OTO commutator 
for the spin versions of the 1CK and 2CK models by 
using numerical time evolving block decimation (TEBD) and exact diagonalization (ED).
In both approaches we actually compute the OTO correlator 
$f(t)=\langle s_z(t) s_z s_z(t) s_z\rangle$, out of which the full commutator $C(t)$ is 
then constructed using Eq.~\eqref{otockondo}. 

Let's discuss first the TEBD approach. The calculation is done in two steps: ($i$) starting from 
a random state we perform an imaginary time evolution to find an approximation for the ground state, 
$\ket {\Psi_{GS}}$, (expressed as a matrix product state (MPS))~\cite{fn2} then, ($ii$) we 
do a real time evolution to compute the OTO correlator $f(t)$. 
We compute $\ket {\Psi_{1}(t)} =e^{iHt} s_z e^{-iHt}s_z\ket {\Psi_{GS}} $ by first acting with 
the $s_z$ operator  (expressed as a Matrix Product Operator (MPO)) on $\ket {\Psi_{GS}}$ followed
by a forward in time evolution i.e. applying the operator $U_{\rightarrow}(t)=e^{-iHt}$, up to time $t$. Then, we act again with the $s_z$ 
and do a backward in time evolution, $U_{\leftarrow}(t)=e^{iHt}$ and get the state $\ket {\Psi_{1}(t)}$. In a similar way, with some interchanged operations, we compute $\ket {\Psi_{2} (t)} =s_z e^{iHt} s_z e^{-iHt}\ket 
{\Psi_{GS}}$. The OTO correlator $f(t)$ can be regarded as the overlap of these two states, $f(t)=\bra 
{\Psi_{2}(t)} \Psi_{1}(t)\rangle$. 

\begin{figure}[h!]
\includegraphics[width=0.95\columnwidth]{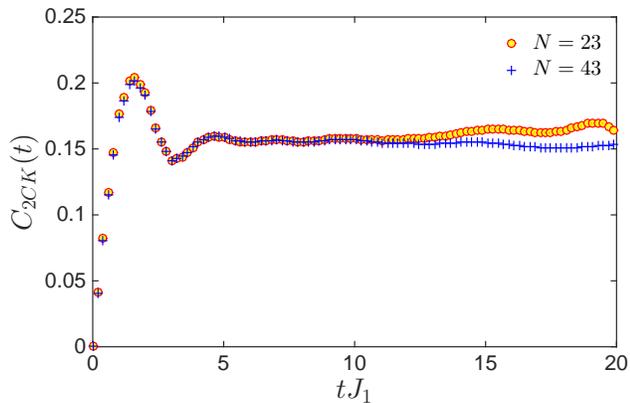}
\caption{ The effect of system size on the OTO commutator $C_{2CK}(t)$ computed using the TEBD algorithm. When the system size is N=23, the break down occurs at $t\simeq 12/J_1$ due to entanglement growth, while  for larger system sizes much later times are accessible. The parameters are fixed to $J'_{R,L}=0.8$.}
\label{fig:scaling}
\end{figure}

The calculation of OTO correlator with the ED approach implies a two step
procedure also: $(i)$
 Finding the ground state within the ED is done by 
representing the  Hamiltonian in Eq.~\eqref{spinonly} numerically as a square matrix with size $2^{N}\times 2^N$ with $N=N_R+N_L+1$ the total number of spins.
As the Hamiltonian ~\eqref{spinonly} is $SU(2)$ invariant, when finding the ground state
we project the  Hamiltonian into the $S=0$ spin sector when $N$ is even, or in the $S=1/2$ sector for $N$ odd to reduce the size of the Hilbert space. 
Then, the ground state is evaluated using the Lanczos algorithm. 
Let us mention
that we have checked our results on the OTO correlator  with the projection to those without any projection, and found perfect agreement.
$(ii)$
The second step is somehow similar to the one described within the TEBD previously with the only 
remark that the time evolutions are evaluated using the Krylov method \cite{Saad-1992}, therefore the explicit evaluation of $e^{-iHt}$ can be avoided.
\begin{figure}[h!]
\includegraphics[width=0.95\columnwidth]{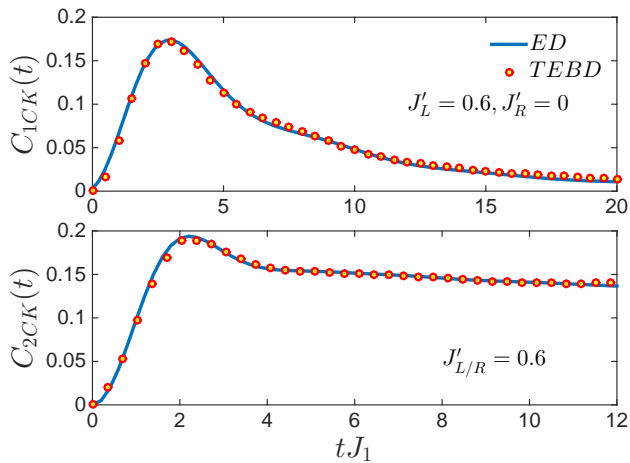}
\caption{ Typical comparison between the exact diagonalization (ED), (solid blue lines)  and time evolving block decimation (TEBD)  (symbols) for the OTO commutator for 1CK (upper panel) and 2CK (lower panel). In both cases the total system size, including the local spin(s) is fixed to N = 21. }
\label{fig:comparison}
\end{figure}
In Fig.~\ref{fig:comparison} we present the results for $C(t)$ computed with the two methods and  
find really good agreement. 
To be able to compare the two approaches we have used chains with lengths
of the order $N\approx 20$, for which ED is still tractable, but  the later time correlator becomes 
affected by finite size effects for $t \gtrsim 15\div 20 /J_1$ in both 1CK and 2CK models. 
To get $C(t)$ at much later times, we need to increase the system size, which is only 
possible within the TEBD approach. 
On the other hand the TEBD suffers from entanglement growth, therefore it also breaks down at some longer times as well.
In Fig.~\ref{fig:scaling} we present the finite size effects for the
 $C_{2CK}(t)$. For chainlengths
$N\approx 40$, later times of the order $t\approx 30/J_1$ are reachable. Furthermore, $C(t\to \infty)$ approaches the infinite value $\sim 0.15$ slightly larger than 1/8 for which the OTO correlator $f(t)$ vanishes.

In the rest of this section we shall discuss the general features characterizing the OTO commutator. 
After an initial sharp peak, 
the OTO commutator saturates to a constant, which is larger than what is expected for spin-1/2 operators in a chaotic system~\cite{maldacena2016,roberts,roberts2016} (i.e. 1/8) for 2CK,
while takes on a rather small value for 1CK. In both cases,  the OTO correlator starts from 1/16 at $t=0$, but  changes sign to take a negative late time value for the 2CK, while
for the 1CK, it re-approaches its initial value, 1/16 at late times.
These highlight the important difference between the 1CK and 2CK cases and the influence of the iQCP on the OTO commutator.

\begin{figure}[h!]
\includegraphics[width=0.8\columnwidth]{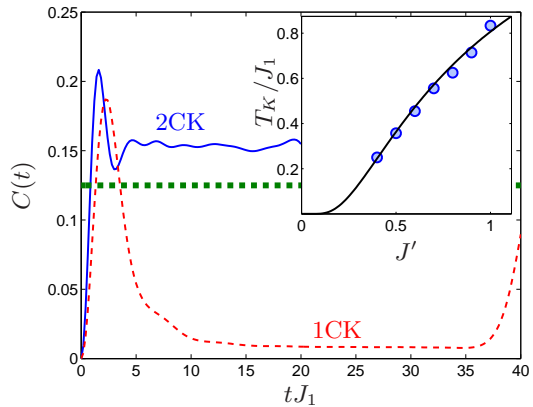}
\caption{The ED data for the OTO commutator for the 1CK (red dashed line) and 2CK (blue solid line) is plotted for $J'_{L}=0.7$, $J'_{R}=0$, 
$N_L=21$ and $J'_{R,L}=0.8$, $N_L=N_R=11$, respectively. Finite size effects are already present for later times.
The thick green dashed  lines denotes the expected late time value when the OTO correlator vanishes in e.g. chaotic systems.
 The inset shows the scaling of the Kondo temperature for 2CK~\cite{mitchell}
 as $T_K=1.82 J_1\exp(-0.82/J')$ (black solid line),  deduced from the position of the initial sharp peak (blue circles).}
\label{kondoed}
\end{figure}

The numerical results 
agree with our analytical findings, even though due to the comparable $J_1$ and $T_K$ in the numerics, we
cannot reach the universal $T_K\ll J_1$ limit, where probably the values at the Emery-Kivelson point would be recovered.
For smaller $J'$, a more enhanced OTO commutator (very close to its maximal value, 1/4) seems to emerge for 2CK from the numerics
together with more suppressed late time value for 1CK,  though
the time evolution cannot be tracked due to finite size effects as we cross over to a Kondo box~\cite{thimm}, when the level spacing becomes larger than the Kondo temperature.
Nevertheless, the distinct behaviour of the local Fermi vs. non-Fermi liquid ground states is clearly observable. 

By associating the Kondo temperature to the sharp peak in $C(t)$, its expected scaling, $T_K\sim \exp(-|c|/J')$ is confirmed in the inset of Fig. \ref{kondoed}.
We suspect that in order to reach the predicted universal values 0 and 1/4 in $C(t\gg 1/T_K)$ for 1CK and 2CK, respectively, 
much bigger system sizes with much smaller Kondo temperature would be required.
The spin-only lattice realization of the two channel Kondo model, Eq.~\eqref{spinonly}, is non-integrable. Therefore, we speculate that this  interferes with non-Fermi liquid nature of the ground state and influences
the OTO commutator: 
for ergodic systems, the above commutator is expected to reach 1/8 at late times~\cite{roberts,roberts2016}. The numerical data indeed indicates that the late time value is in between 1/8 and 1/4, expected for chaotic systems and for the
integrable 2CK, respectively.

\section{Conclusion}

We have investigated the OTO correlator around an impurity quantum phase transition in the Kondo model analytically and numerically.
By making use of its mapping onto the Majorana resonant level model, we identify a temperature independent OTO commutator for the impurity spin in the two channel case, 
in contrast to its simple
commutator counterpart.
Though this excludes the Lyapunov growth,  the late time value of the OTO commutator reveals salient features of the underlying models.
For the single channel realization, the ground state is a Fermi liquid and the OTO commutator vanishes at late times (or at most preserves a tiny finite value) and the quantum butterfly effect is absent.
 For the two channel model, on the other hand, the competition between the two channels in trying to screen 
 the impurity spin  produces a local non-Fermi liquid ground state. 
The OTO commutator exhibits a sizeable late time value in spite of the integrability of the model, representing an enhanced butterfly effect.
This is a direct consequence of fractionalization as a Majorana mode from the impurity decouples completely.

\section*{acknowledgments}

This research is supported by the National Research, Development and Innovation Office - NKFIH  K105149, K108676, SNN118028 and K119442. CMP was supported by Romanian UEFISCDI, project number PN-III-P4-ID-PCE-2016-0032.

\bibliographystyle{apsrev}
\bibliography{wboson1}

\begin{thebibliography}{10}
\bibitem{fn3}
\bibinfo{note}{In general for any $A$, $B$, $C$ and $D$ linear in the Majorana
  operators we can use the decoupling $\langle ABCD\rangle=\langle AB\rangle
  \langle CD\rangle-\langle AC\rangle \langle BD\rangle+\langle AD\rangle
  \langle BC\rangle$, where the wavefunction is a Slater determinant.}
\bibitem{fn2}
\bibinfo{note}{We have compared the ground state energy obtained in this way
  against the one obtained using the density matrix renormalization group
  (DMRG) and found perfect agreement}.
\end{thebibliography}

\end{document}